\documentclass[letter]{aa} % for the letters 
\usepackage{graphicx}
\usepackage{txfonts}
\usepackage{mathrsfs}  
\let\mathcal=\mathscr
\usepackage[colorlinks,allcolors=blue]{hyperref}
\begin{document}

   \title{A prediction about the age of thick discs as a function of the stellar mass of the host galaxy}

   \author{S.~Comer\'on
          \inst{1,2}}

    \institute{Departamento de Astrof\'isica, Universidad de La Laguna, E-38200, La Laguna, Tenerife, Spain\\ \email{seb.comeron@gmail.com}
              \and Instituto de Astrof\'isica de Canarias E-38205, La Laguna, Tenerife, Spain
              }

% \abstract{}{}{}{}{} 
% 5 {} token are mandatory
 
  \abstract{One of the suggested thick disc formation mechanisms is that they were born quickly and in situ from a turbulent clumpy disc. Subsequently, thin discs formed slowly within them from leftovers of the turbulent phase and from material accreted through cold flows and minor mergers. In this letter, I propose an observational test to verify this hypothesis. By combining thick disc and total stellar masses of edge-on galaxies with galaxy stellar mass functions calculated in the redshift range of $z\leq3.0$, I derived a positive correlation between the age of the youngest stars in thick discs and the stellar mass of the host galaxy; galaxies with a present-day stellar mass of $\mathcal{M}_\star(z=0)<10^{10}\,\mathcal{M}_\odot$ have thick disc stars as young as $4-6$\,Gyr, whereas the youngest stars in the thick discs of Milky-Way-like galaxies are $\sim10$\,Gyr old. I tested this prediction against the scarcely available thick disc age estimates, all of them are from galaxies with $\mathcal{M}_\star(z=0)\gtrsim10^{10}\,\mathcal{M}_\odot$, and I find that field spiral galaxies seem to follow the expectation. On the other hand, my derivation predicts ages that are too low for the thick discs in lenticular galaxies, indicating a fast early evolution for S0 galaxies. I propose the idea of conclusively testing whether thick discs formed quickly and in situ by obtaining the ages of thick discs in field galaxies with masses of $\mathcal{M}_\star(z=0)\sim10^{9.5}\,\mathcal{M}_\odot$ and by checking whether they contain $\sim5$\,Gyr-old stars.}

   \keywords{Galaxies: evolution -- Galaxies: spiral -- Galaxies: structure
               }

   \maketitle
%
%-------------------------------------------------------------------

\section{Introduction}

Thick discs in galaxies are the low surface brightness and large scale-height counterparts of the canonical thin discs. Thick disc stars are older than those of the thin discs (e.g.~\citealt{Bensby2005} for the Milky Way, and \citealt{Yoachim2008, Comeron2015, Pinna2019} for other galaxies) and they comprise at least several tens of percent of the baryons in discs (e.g.~\citealt{Reid2005, Fuhrmann2008, Fuhrmann2012, Snaith2014} for the Milky Way, and \citealt{Yoachim2006, Comeron2011, Comeron2018} for other galaxies). Hence, thick discs are key to understand how early galaxies evolved.

The origin of thick discs is controversial. The proposed formation mechanisms cover the four corners of the diagram that summarises the processes of galactic evolution in \citet{Kormendy2004}. The position of a mechanism in this diagram is characterised by two parameters: whether it is internally- or environment-driven, and whether the process is fast (of the order of a dynamical timescale) or secular. In this framework, the possible thick disc formation mechanisms are as follows: 1) external fast processes, where the thick disc is a consequence of the merger of gas-rich galaxies during the initial galaxy assembly process \citep{Brook2004, Martig2014}, 2) internal fast processes where the thick disc is born thick due to the turbulent and clumpy nature of the first discs \citep{Elmegreen2006, Bournaud2009, Comeron2014}, 3) external secular processes where the thick disc is made of stars stripped from infalling satellites \citep{Abadi2003} and by the disc dynamical heating caused by satellites \citep{Qu2011}, and 4) internal secular processes, where the thick disc is caused by dynamical heating due to disc overdensities \citep{Villumsen1985} and/or the radial migration of stars \citep{Schoenrich2009, Roskar2013}.

Several of the abovementioned mechanisms might be acting concurrently. For example, simulations have shown that the combination of internally and externally driven secular heating can produce realistic thick discs \citep{Minchev2015, GarciadelaCruz2020}. Another example is the fact that the thick discs in three S0 galaxies in the Fornax cluster show traces of a significant minority of accreted stars on top of a population that was formed in situ \citep{Pinna2019, Pinna2019a}. It is also possible that galaxies with different masses have different thick disc formation paths \citep{Comeron2012}.

In this letter, I present a prediction about the age of the youngest stars of thick discs as a function of the stellar mass of the host galaxy if the main formation mechanism is internal and fast. The confrontation between this prediction and forthcoming spectroscopically-derived thick disc ages will make it possible to confirm or reject the scenario put to the test.

I assume that the formation of thick disc stars is completed before the thin disc starts forming \citep{Elmegreen2006, Bournaud2009} and that the thick disc and the classical bulge form simultaneously \citep{Comeron2014}. In this framework, the first discs were clumpy and with a large star formation rate, roughly ten times larger than that in the present day. Because the star formation scale height scales with the specific star formation rate \citep{Comeron2014}, these discs were thick. Due to dynamical friction, some of the clumps spiralled inwards and formed bulges. The thick disc and the classical bulge are collectively called the dynamically hot components.

\section{Derivation}

\label{derivation}

Here I derive the age of the youngest stars in thick discs (and classical bulges) as a function of the host stellar mass for thick discs which form quickly and in situ. Given a galaxy with a present-day stellar mass $\mathcal{M}_{\star}(z=0)$, two elements are required to do so: 1) the mass of the dynamically hot components, $\mathcal{M}_{\rm h}$, and 2) the evolution with redshift of the total stellar mass, $\mathcal{M}_{\star}(z)$.

\begin{figure}
   \centering
   \includegraphics[scale=0.35]{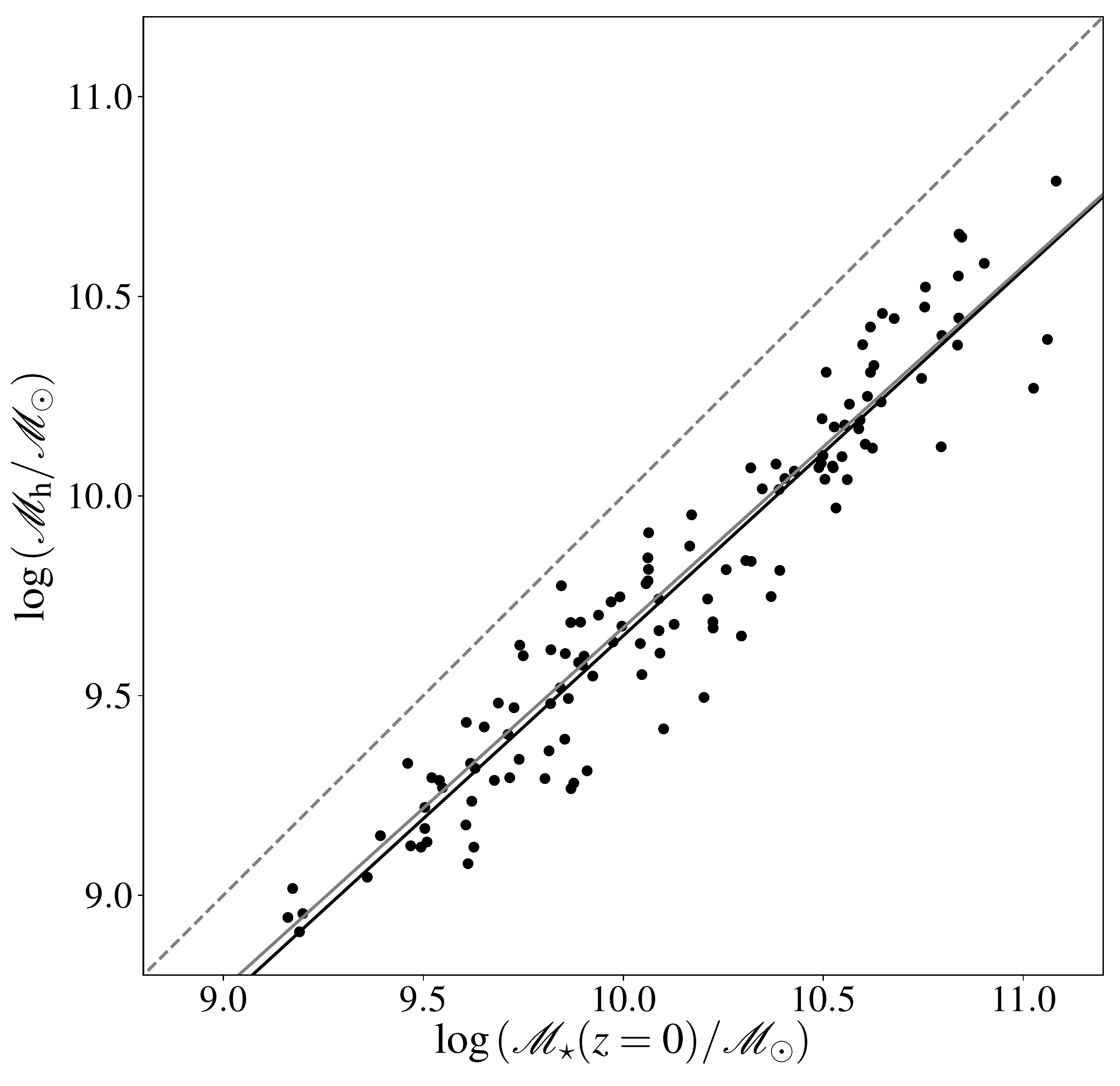}
   \caption{\label{scatter}{Scatter plot of the sum of the masses of the thick disc and the classical bulge (dynamically hot components; $\mathcal{M}_{\rm h}$) versus the total stellar mass of the galaxy, $\mathcal{M}_{\star}(z=0)$, for the 124 galaxies with a fitted thick disc in \citet{Comeron2018}. The black continuous line indicates the linear fit to all the data, whereas the continuous grey line indicates the fit to the data points corresponding to galaxies in the Virgo and Fornax clusters only. The dashed grey line indicates a one-to-one relation.}}
\end{figure}

The first point can be addressed by resorting to the decompositions of {\it Spitzer} Survey of Stellar Structure in Galaxies \citep[S$^4$G;][]{Sheth2010} edge-on galaxies by \citet{Comeron2018}, which provide the masses of the thin discs, the thick discs, and the classical bulges. The risk of confusion between pseudobulges and classical bulges was minimised by assigning the light of flattened components to the disc (usually the thin disc). Fig.~\ref{scatter} displays the hot component mass, $\rm{log}\,(\mathcal{M}_{\rm h}/\mathcal{M}_\odot)$, as a function of the total stellar mass, $\rm{log}\,(\mathcal{M}_{\star}(z=0)/\mathcal{M}_\odot)$, for the 124 galaxies with a fitted thick disc in \citet{Comeron2018}. A linear regression to the scatter plot gives
\begin{equation}
 \label{fit}
 {\rm log}\,(\mathcal{M}_{\rm h}/\mathcal{M}_\odot)=0.489+0.9161\,{\rm log}\,(\mathcal{M}_{\star}(z=0)/\mathcal{M}_\odot).
\end{equation}

To find how the stellar mass of galaxies evolves with time, I followed the method proposed by \citet{Dokkum2013}. For a given stellar mass at present, $\mathcal{M}_{\star}(z=0)$, I find the progenitors at a given redshift, $z$, by assuming that the comoving density does not vary with time. This approach is valid as long as galaxies conserve their mass rank order, which is true within 0.15\,dex in the simulations by \citet{Leja2013}. In practice, this is done by integrating galaxy stellar mass functions at different redshifts, $\phi(\mathcal{M},z)$, to find the mass, $\mathcal{M}_{\star}(z)$, above which the cumulative comoving density is equal to that for the present-day stellar mass $\mathcal{M}_{\star}(z=0)$
\begin{eqnarray}
\label{integral}
 \rho(\mathcal{M}>\mathcal{M}_{\star}(z=0))=\int_{\mathcal{M}_{\star}(z=0)}^{\infty}\phi(\mathcal{M},z=0)\,d\mathcal{M}=\\ \nonumber
 =\int_{\mathcal{M}_{\star}(z)}^{\infty}\phi(\mathcal{M},z)\,d\mathcal{M}=\rho(\mathcal{M}>\mathcal{M}_{\star}(z)).
\end{eqnarray}

\begin{figure}
   \centering
   \includegraphics[scale=0.35]{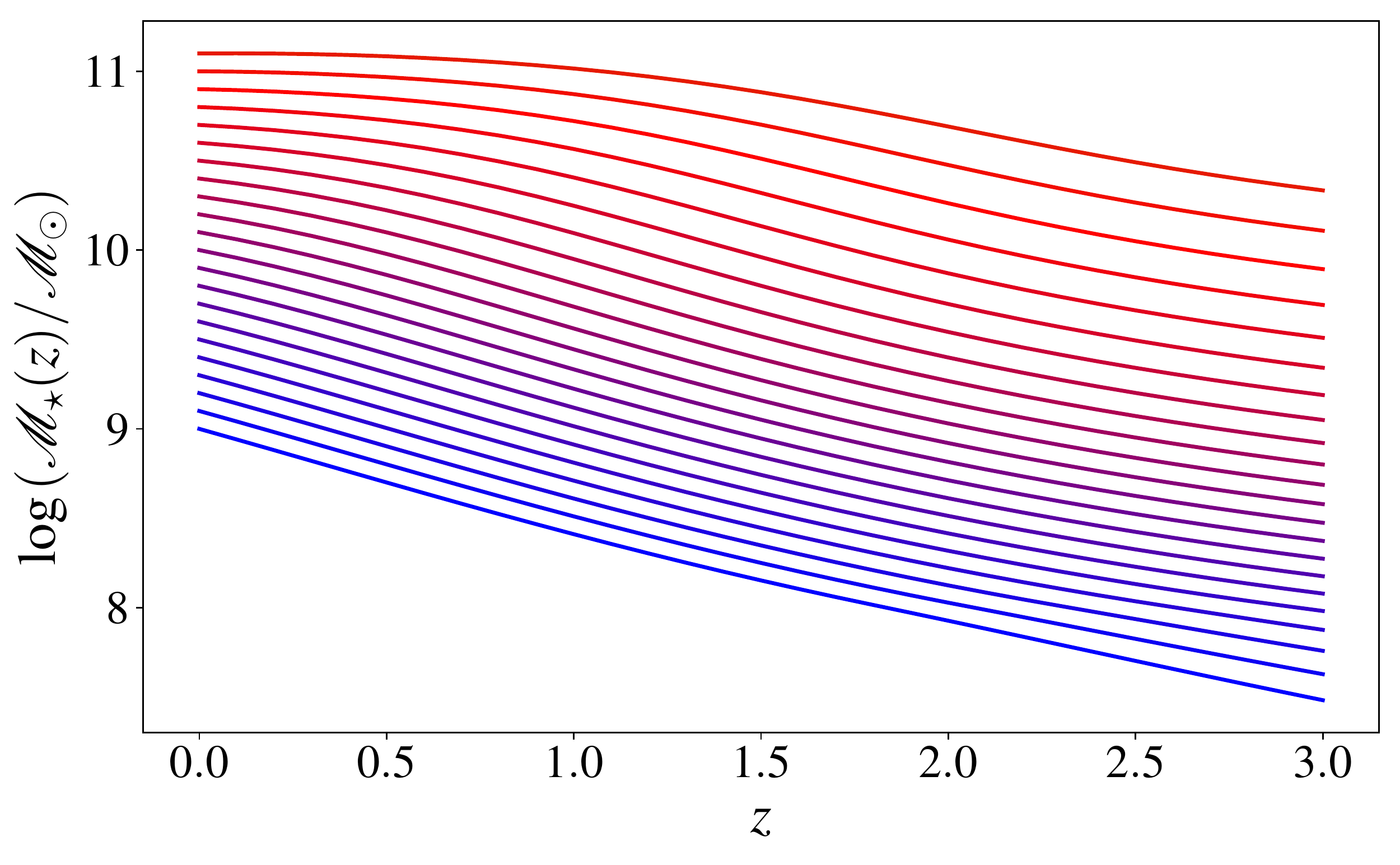}
   \caption{\label{growth}{Stellar mass as a function of redshift for galaxies with present-day stellar masses between $10^9$ and $10^{11.2}\,\mathcal{M}_\odot$ (in 0.1\,dex intervals) calculated using the continuity model from \citet{Leja2020}.}}
\end{figure}

\citet{Leja2020} compiled data from galaxies in the redshift range of $0.2\leq z\leq3.0$ and constructed a continuity model to describe $\phi(\mathcal{M},z)$ at an arbitrary redshift within the studied range. Their model was built using all the masses and redshifts of the galaxies in their sample simultaneously, without using any sort of binning in either mass or redshift. I computed the mass functions with a \texttt{python} code provided by \citet{Leja2020} and used the mass functions and Eq.~\ref{integral} to derive the stellar mass evolution of galaxies with $\mathcal{M}_{\star}(z=0)$ between $10^9$ and $10^{
11.2}\,\mathcal{M}_\odot$. The evolution at $z<0.2$ was calculated with extrapolated stellar mass functions. The resulting stellar mass evolution estimates are presented in Fig.~\ref{growth}.

\begin{figure}
   \centering
   \includegraphics[scale=0.35]{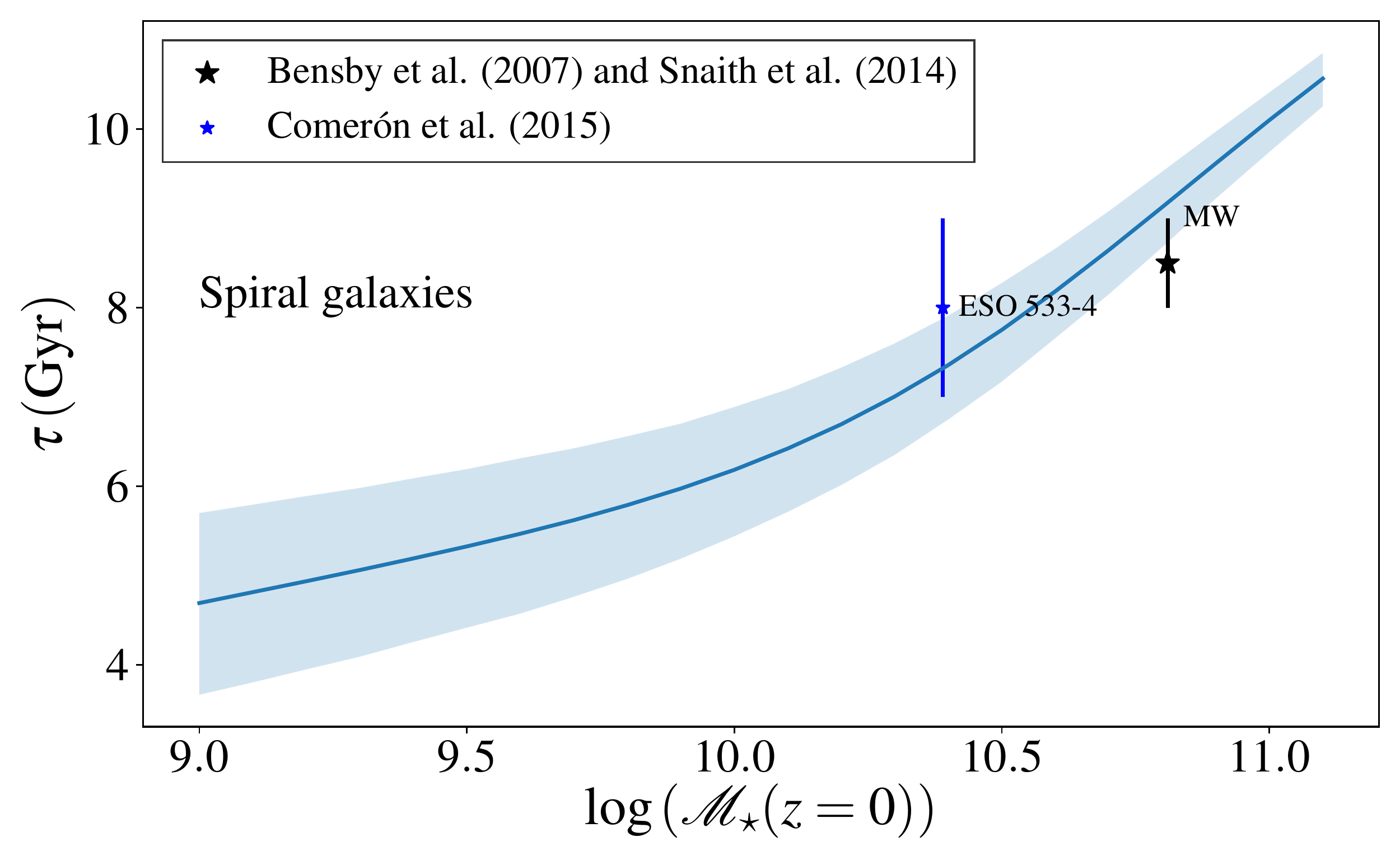}
   \caption{\label{age_sp}{Predicted age of the youngest stars in thick discs as a function of the total present-day stellar mass of the host. The error bands correspond to one-$\sigma$ confidence intervals. The symbols correspond to spiral galaxies with thick disc stellar ages and masses reported in the literature.}}
\end{figure}

Under the assumption that the thick disc and the classical bulge form before the thin disc, Eq.~\ref{fit} and the information in Fig.~\ref{growth} can be combined to obtain, for each present-day stellar mass, $\mathcal{M}_{\star}(z=0)$, the redshift when the thick disc formation ended, that is the redshift for which
\begin{equation}
 \mathcal{M}_{\rm h}=\mathcal{M}_{\star}(z).
\end{equation}
These redshifts can be transformed into an age, $\tau$, so they can be compared with the observed thick disc ages. For consistency with \citet{Leja2020}, I assume the WMAP9 cosmology \citep{Hinshaw2013}. The resulting age of the youngest stars in thick discs as a function of the host stellar mass is shown in Figs.~\ref{age_sp}.

\begin{figure}
   \centering
   \includegraphics[scale=0.35]{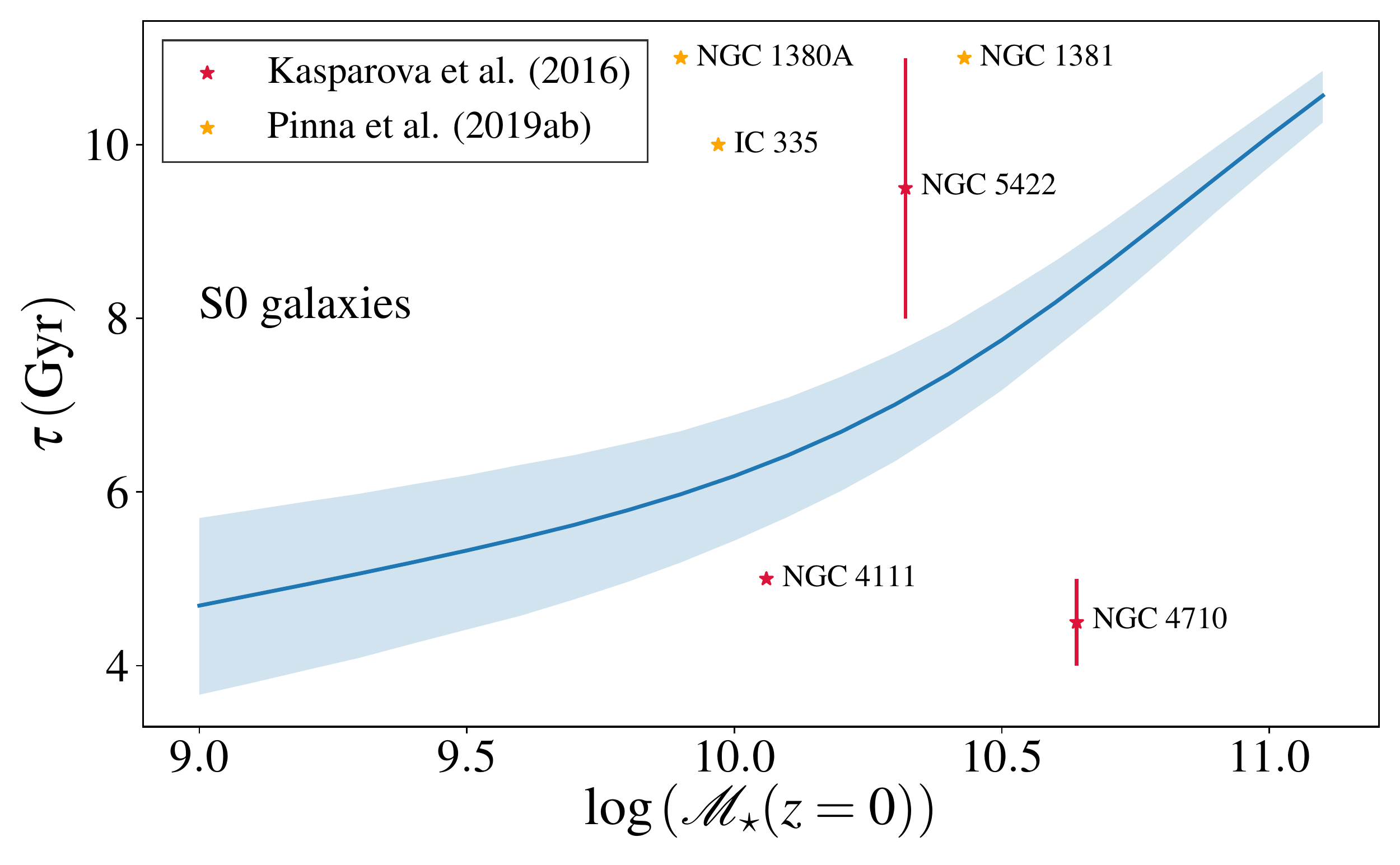}
   \caption{\label{age_s0}{Same as Fig.~\ref{age_sp}, but with data points corresponding to S0 galaxies.}}
\end{figure}

I estimated confidence intervals with a Monte-Carlo approach by redoing the linear fit in Eq.~\ref{fit} from bootstrapped data points drawn from Fig.~\ref{scatter} and combining the new fit with a new version of the curves in Fig.~\ref{growth} calculated from the \citet{Leja2020} model with the parameters randomised assuming one-$\sigma$ Gaussian uncertainties. I carried out 1000 repetitions to estimate the one-$\sigma$ uncertainty bands in the figures.

Figure~\ref{age_sp} shows that massive galaxies are expected to have very old thick discs with ages around 10\,Gyr. Lower-mass galaxies, notably those with stellar masses below $10^{10}\,\mathcal{M}_\odot$, are expected to have thick discs that stopped forming stars much later, between 4 and 6\,Gyr ago. The curve shown in Fig.~\ref{age_sp} is closely matched by a quadratic function in the mass range of interest
\begin{eqnarray}
\label{prediction}
 \tau\,({\rm in\ Gyr})=\\ \nonumber
 =104-22.15\,\rm{log}\,(\mathcal{M}_{\star}(z=0))+1.237\,({\rm log}\,(\mathcal{M}_{\star}(z=0)))^2.
\end{eqnarray}

\section{Comparison with observations}

\label{observations}

In this section, the prediction that thick discs are older in the most massive galaxies is confronted with thick disc ages that were obtained spectroscopically for a few galaxies with thick disc stellar masses derived in \citet{Comeron2018}. Strictly speaking, the prediction concerns the age of the youngest stars in the thick disc. However, from an observational perspective, the difference is small because the luminosity of a galaxy component is dominated by its youngest stars.

The observational data and thick disc ages have been obtained by different groups. Hence, the data are heterogeneous, both in their acquisition and processing, as explained below. The recovery of star formation histories from spectra is an ill-posed inverse problem, which might result in large uncertainties in the estimated ages. Furthermore, bursty star formation is smoothed by regularisation. As a consequence, if the thick discs formed in a short burst, the recovered star formation history would be an artificially broadened peak causing an underestimation of the age of the youngest thick disc stars.

I find that the two non-lenticular galaxies, the Milky Way and ESO~533-4, approximately fall in the relation derived in Sect.~\ref{derivation} (Fig.~\ref{age_sp}). On the other hand, lenticular galaxies do not follow the relation (Fig.~\ref{age_s0}).

{\it ESO~533-4 \citep{Comeron2015}:} The data were obtained with the optical integral field spectrograph (IFS) VIMOS at the VLT. The thick disc age was estimated with pPXF \citep{Cappellari2004}. In order to account for a possible excessive regularisation-induced smoothing, I estimated the age of the newest stars in the thick disc to be $7-9$\,Gyr by finding the interval between which 90\% and 99\% of thick disc stars were formed according to the fit.

{\it NGC~4111, NGC~4710, and NGC~5422 \citep{Kasparova2016}:} The data were obtained with the single-slit spectrographs SCORPIO and SCORPIO-2 at BTA. The ages of the thick discs were estimated with {\sc nbursts} \citep{Chilingarian2007, Chilingarian2007a}. This yielded 5\,Gyr for NGC~4111, $4-5$\,Gyr for NGC~4710, and $8-11$\,Gyr for NGC~5422. In the latter two galaxies, however, the slit was placed in regions where the light is dominated by the thin disc according to \citet{Comeron2018}. In the case of NGC~4111, the galaxy is thick disc-dominated everywhere at $3.6\,{\mu}{\rm m}$ \citep{Comeron2018}, but the slit was placed only 400\,pc above the midplane, in a region where the thin disc has a significant contribution at $3.6\,{\mu}{\rm m}$, which is probably even larger in the optical. Therefore, the estimated thick disc ages for these galaxies might be significantly biased downwards by the thin disc.

{\it IC~335, NGC~1380A, and NGC1381 \citep{Pinna2019, Pinna2019a}:} The observations of these S0 galaxies in the Fornax cluster were made with the IFS MUSE at the VLT. The authors obtained thick disc star formation histories with pPXF and found that the thick discs have a small fraction of accreted stars that are younger than the in situ population. The accreted stars are distinguishable from those born in situ because of their high [Mg/Fe]. As this letter is about in situ thick discs, the age that has been adopted is that of the youngest age bin that is not [Mg/Fe]-enhanced (10\,Gyr for IC~335, 11\,Gyr for NGC~1380A, and 11\,Gyr for NGC~1381). If some residual in situ thick disc star formation remained by then, the age of the youngest stars in the thick disc would be overestimated.

Data for the Milky Way are also included. I used an age range of $8-9$\,Gyr \citep{Bensby2007} for the youngest thick disc stars, which is compatible with an age of 9\,Gyr found by \citet{Snaith2014}. I also used a stellar mass of $\mathcal{M}_\star(z=0)=5\times10^{10}\,\mathcal{M}_\odot$ \citep{McMillan2011}.

\section{Discussion: Potential different evolution for the thick discs in spiral and S0 galaxies}

The derivation in Sect.~\ref{derivation} shows that if thick discs form quickly and in situ, those in low-mass galaxies form slower. This is the consequence of two concurrent factors: 1) the smaller the total stellar mass of a galaxy, the larger the relative thick disc mass \citep{Yoachim2006, Comeron2011, Comeron2012, Comeron2014, Comeron2018}; and 2) downsizing, which is the process by which large-mass galaxies form faster than small-mass ones \citep{Cowie1996}. These two factors are demonstrated in Figs.~\ref{fit} and \ref{growth}, respectively.

Based on the scarce observational evidence, the predicted trend between the age of the youngest stars in a thick disc and the total stellar mass is not followed by S0 galaxies. On the other hand, the two spiral galaxies for which we have data seem to follow it. Assuming that the literature data are representative of the whole population, it is important to determine what the implications are.

Two of the three lenticular galaxies in \citet{Kasparova2016} have thick disc ages that fall well below the expectation (Fig.~\ref{age_s0}), although most likely, they are measuring a population with a significant thin disc contribution. On the other hand, some lenticular galaxies -- including the three galaxies in the Fornax cluster -- have ages that are above those predicted. This might be due to a difference in evolutionary paths between isolated galaxies and those in clusters. \citet{Silchenko2012} and \citet{Comeron2016} propose that, for at least some S0 galaxies, the evolution when entering a cluster is greatly accelerated (perhaps almost as in a monolithic collapse). Regarding the galaxies discussed in Sect.~\ref{observations}, this is clearly the case for NGC~1381, where the thin and the thick discs show no age difference \citep{Pinna2019a}. Because the thin disc is more metal-rich than the thick disc, it is likely that the thin disc is slightly younger, but the difference in age cannot be resolved \citep[see the case of ESO~243-49 in][]{Comeron2016}. On the other hand, IC~335 and NGC~1380A have thin discs whose mean age is noticeably younger than the thick disc, perhaps indicating a `traditional' evolution where star formation is not quenched by the environment, or indicating that the galaxy has been rejuvenated by accreting fresh material, as suggested by \citet{Katkov2019}.

The possibility of two evolutionary paths distinguishing galaxies in the field and at least some lenticular galaxies in clusters is intriguing. Furthermore, it could undermine some of the premises under which this study was done: 1) that all the galaxies fall into a single ${\rm log}\,\mathcal{M}_{\rm h}$ versus~${\rm log}\,\mathcal{M}_{\star}(z=0)$ relation, and 2) that the evolution of the stellar mass function does not depend on the galaxy type. Point 1) was tested by redoing the fit in Fig.~\ref{fit} with the 31 galaxies belonging to either the Virgo or the Fornax clusters (continuous grey line). The differences between the fit including cluster galaxies and that with all galaxies are insignificant. This indicates that the differences between spiral and lenticular galaxies are probably related to different mass growth rates.

If the mass evolution differs significantly between cluster and field galaxies, the relationship between the age of the youngest stars in the thick disc and the stellar mass $\mathcal{M}_{\star}(z=0)$ presented in Eq.~\ref{prediction} is a weighted average of two relations. The first one is a relation for cluster disc galaxies or, at least, for some cluster lenticulars. This relation would be much shallower than that in Eq.~\ref{prediction}, and probably nearly flat at $\tau\sim10\,{\rm Gyr}$ for the range of stellar masses sampled by the observations in Sect.~\ref{observations}, that is $\mathcal{M}_{\star}(z=0)\gtrsim10^{10}\,\mathcal{M}_\odot$. The second relation is for field disc galaxies. This relation would fall below that presented in Eq.~\ref{prediction} in the mass range of $\mathcal{M}_{\star}(z=0)\gtrsim10^{10}\,\mathcal{M}_\odot$. Because of the many assumptions made in the derivation of Eq.~\ref{prediction}, and the fact that field and cluster galaxies probably have significantly different evolutionary paths, a scatter larger than that presented by the error bands in Figs.~\ref{age_sp} is expected.

If the thick discs hosted in field galaxies with stellar masses of $\mathcal{M}_\star(z=0)\sim10^{9.5}\,\mathcal{M}_\odot$ are significantly younger than thick discs in Milky-Way-like galaxies (e.g.~ages of $\sim5$\,Gyr versus $\sim10$\,Gyr for the Milky Way), it would be a good indication that dynamically hot components form before the thin disc does. Additional observations with world-class spectrographs such as MUSE could indeed provide such constraints.

\section{Conclusions}

In order to shed light on the origin of thick discs, I propose a test to verify one of the proposed formation mechanisms. The test concerns the hypothesis that thick discs formed quickly and in situ from turbulent clumpy discs before the thin disc formed within them \citep{Elmegreen2006, Bournaud2009, Comeron2014}.

By combining the thick disc and the total stellar masses of edge-on galaxies in the S$^4$G from \citet{Comeron2018} and a model allowing me to calculate galaxy stellar mass functions in the redshift range of $0.2\leq z\leq3.0$ \citep{Leja2020}, I derived a relationship between the age of the youngest stars in the thick discs and total stellar mass of the host. I thus predict that the age of the youngest thick disc stars increases with the host stellar mass: as young as $4-6$\,Gyr in galaxies with a total stellar mass below $\mathcal{M}_\star(z=0)\sim10^{10}$, versus $\sim10$\,Gyr in Milky-Way-like galaxies.

The scarce available observational data indicate that spiral field galaxies (including the Milky Way) seem to follow the expectation. However, lenticular galaxies in clusters have thick disc stars that are older than predicted, perhaps indicating a very fast early evolution as suggested by \citet{Silchenko2012} and \citet{Comeron2016}.

To strengthen the tentative evidence that field galaxies formed their thick discs in situ and quickly at high redshift, further observations are required, especially for the unexplored mass range of $\mathcal{M}_\star(z=0)<10^{10}\,\mathcal{M}_\odot$. We might expect to find that galaxies with a total stellar mass of $\mathcal{M}_\star(z=0)\sim10^{9.5}\,\mathcal{M}_\odot$ have thick disc stars as young as $\sim5$\,Gyr.

\begin{acknowledgements} 
I thank the anonymous referee for comments that have helped to improve this letter. I thank Dr.~Sim\'on D\'iaz-Garc\'ia for useful discussions and Prof.~Johan H.~Knapen for revising the manuscript. I acknowledge support from the State Research Agency (AEI-MCINN) of the Spanish Ministry of Science and Innovation under the grant `The structure and evolution of galaxies and their central regions' with reference PID2019-105602GB-I00/10.13039/501100011033.

\end{acknowledgements}

\bibliographystyle{aa} % style aa.bst
\bibliography{bibliography.bib} % your references Yourfile.bib

\end{document}